\documentclass[12pt]{article}
\usepackage{psfrag}

\newcommand{\apj}{Astrophys. J}
\newcommand{\apjs}{Astrophys. J. Suppl.}
\newcommand{\prd}{Phys. Rev. D}
\newcommand{\aanda}{Astron. Astrophys}
\begin{document}
\def\stackunder#1#2{\mathrel{\mathop{#2}\limits_{#1}}}





\title{An analytic approach to the late ISW effect in a $\Lambda$ dominated universe}

\author{\small Viktor Czinner, M\'aty\'as Vas\'uth and \'Arp\'ad Luk\'acs\\
\small KFKI Research Institute for Particle and Nuclear Physics,\\
\small Budapest 114, P.O.Box 49, H-1525 Hungary\\ 
\small czinner@rmki.kfki.hu, vasuth@rmki.kfki.hu, arpi@rmki.kfki.hu}
\date{}
\maketitle


\begin{abstract}
The integrated Sachs-Wolfe (ISW) effect in a $\Lambda$ dominated universe can be an important factor in the evolution of cosmic microwave background fluctuations. With the inclusion of cosmological constant we present the complete analytic solution of the covariant linear perturbations of the Einstein equations in the Newtonian gauge, in the case of a spatially flat ($K=0$), Friedman-Robertson-Walker (FRW) universe, filled with pressureless ideal fluid. We use the analytic time dependence of the perturbation potentials to derive the anisotropy power spectrum of the late ISW effect. We choose the scale invariant Harrison-Zeldovich spectrum for obtaining the $C_{\ell}$ momenta and show the agreement of our results with earlier, numeric calculations.      
\end{abstract}
\vskip 12pt

\noindent
Keywords {\it Relativistic Cosmology, Background radiations}

\section{Introduction}

It is well known that several effects give contribution to the fluctuations of the cosmic microwave background radiation (CMBR). It is usual to divide these contributions to primary and secondary anisotropies. The dominant part of primary anisotropies comes from the Sachs-Wolfe effect \cite{SW}. In a $\Lambda$ or a curvature dominated universe the gravitational potentials of the perturbations
are varying in time, giving rise to the appearance of the ISW effect. In linear theory for $\Omega_0=1$, flat universe, the potentials are constants so this appearance does not occur. 

The ISW effect in a $\Lambda$ dominated universe has been studied first by \cite{Starobinsky} and 
recently in more detailes, as different measurements e.g. {\it Wilkinson Microwave Anisotropy Probe} 
(WMAP) \cite{Spergel}, Ia type supernovae observations \cite{Tonry} and the {\it Sloan Digital Sky Survey} (SDSS) \cite{Tegmark} imply the accelerating expansion of the universe. In the light of this unexpected result the understanding of the properties of $\Lambda$ models becomes more and more important. 
In a thorough overview \cite{HS} Hu and Sugiyama summarize the CMBR anisotropies concerning many different scenarios. In \cite{HW} Hu and White investigate the CMBR anisotropies in the weakly coupled regime (i.e. when the photons travel freely from the last scattering surface to present) including the $\Lambda$ dominated epoch. These works give an accurate picture of present understanding of the microwave background fluctuations using both analytic and numerical methods, although none of them have the general analytic expression for the perturbation potentials in the presence of cosmological constant.   

In our paper, following the treatment of earlier works we calculate the late ISW effect in a $\Lambda$ dominated universe. In Sec. 2, we present the metric with linear perturbations for a flat, homogeneous and isotropic universe. Using the analytic results of \cite{cvlp} we obtain the complete solution of the perturbed problem in the Newtonian gauge for scalar perturbations. In Sec. 3, we calculate the $C_\ell$ momenta and the power spectrum of the anisotropies for the late ISW effect and show that 
our results are in good agreement with earlier full numeric calculations.

\section{The Newtonian potentials of the perturbations}

The unperturbed metric in a homogeneous and isotropic universe is the FRW metric 
\begin{eqnarray} \label{ds2}
ds^2 = -dt^2+a^{2}(t)\gamma_{\mu\nu}dx^\mu dx^\nu \ ,
\end{eqnarray}
with the three metric $\gamma_{\mu\nu}$ of a space with constant spatial curvature $K$. The scale factor $a(t)$ can be expressed in the case of a flat ($K=0$) space and pressureless ($p=0$) matter source as~\cite{Stephani}
\begin{eqnarray}  \label{backgr}
a=a_0 {\rm sinh}^{2/3}(Ct+C_0)\ ,\quad  \rho a^3={\cal C}_M \ , \quad 
a_0=\left(\frac{{\cal C}_M}{\Lambda}\right)^{1/3}\ ,\quad 
C=\frac{\sqrt{3\Lambda}}{2} \ ,
\end{eqnarray}
where, and throughout this paper we use units in which the gravitational constant $G=1/8\pi$, the speed of light $c=1$ and we set $C_0=0$. 

The linear perturbations of the metric tensor for a spatially flat, FRW universe have the general form 
\begin{eqnarray} \label{metric}
g_{00} &=& -(1+2A) \ , \nonumber\\
g_{0\alpha} &=& -a B_\alpha \ , \\
g_{\alpha\beta} &=& a^{2} \left[\left(1+2H_L\right)\gamma_{\alpha\beta} 
 + 2H_{T\alpha\beta}\right] \ ,\nonumber
\end{eqnarray}
where the functions $A$, $H_L$, $B_\alpha$ and $H_{T\alpha\beta}$ give a complete representation
of the metric and $H_{T\alpha\beta}$ is a $3\times 3$ trace-free tensor. In Bardeen's covariant linear perturbation formalism the Einstein equations can be decoupled into a set of ordinary differential equations by employing scalar, vector and tensor eigenmodes of the Laplacian operator, which form a complete set. In the following we will represent the perturbation quantities with the amplitudes of the eigenmodes corresponding to the $k$th wavenumber, see \cite{Bardeen}.

In covariant linear formalism three independent, gauge invariant quantities can be constructed, namely $\Psi$ and $\Phi$ from the metric tensor amplitudes and $V$ from the matter velocity \cite{Bardeen}. These gauge invariant quantities are 
\begin{eqnarray}\label{giq}
\Psi&=&A+\frac{1}{k}\frac{d}{dt}(aB)-\frac{1}{k^2}\frac{d}{dt}\left(a^2\dot{H}_T\right)\ ,\nonumber\\
\Phi&=&H_L+\frac{1}{3}H_T+\frac{\dot aB}{k}-\frac{1}{k^2}a\dot a \dot{H}_T\ ,\\
V&=&v-\frac{1}{k}a\dot{H}_T\ ,\nonumber
\end{eqnarray}
where a dot denotes the derivative with respect to the comoving time coordinate $t$ and $v$ is the velocity perturbation of the matter field. We use the complete first order solution given in comoving coordinates \cite{cvlp}, to obtain the metric tensor amplitudes in the Newtonian (or longitudinal) gauge. This transformation can be done with the help of the gauge invariant quantities in Eq. (\ref{giq}) using the Newtonian gauge conditions $H^N_T=B^N=0$, where the superscript $N$ refers the Newtonian gauge. Performing the calculations yields the solution  
\begin{eqnarray}
\Psi \equiv A^N&=&\frac{C\cosh(Ct)\left\{3C_MB_0(k)- 2Ca_0^6\left[K_1(k) - K_2(k)I(t)\right]\right\}}{3a_0k^2C_M\sinh^{5/3}(Ct)} , \\
\Phi \equiv H^N_L &=& -\Psi \ ,\\
V \equiv v^N&=&\frac{1}{\sinh^{4/3}(Ct)}\left\{\frac{Ca_0^4}{kC_M}\left[K_1(k) - K_2(k)I(t)\right]-\frac{3B_0(k)}{2ka_0^2}\right\} \nonumber\\
&+&\frac{2^{1/3}Ca_0^4}{kC_M}\frac{\sinh^{1/3}(Ct)}{\cosh(Ct)}K_2(k) \ .
\end{eqnarray}
The density perturbation from the field equations also can be obtained immediately as
\begin{eqnarray}\label{drho}
\delta\rho^N&=&\frac{\cosh(Ct)}{\sinh^{3}(Ct)}\left\{\frac{4a_0^3C^2}{3C_M}[K_1(k)-K_2(k)I(t)]-\frac{2C}{a_0^3}B_0(k)\right\}\nonumber \\ 
&+&\frac{2}{3\sinh^{4/3}(Ct)}\left[\frac{k^2}{a_0^2}H_0(k)+\frac{2^{4/3}C^2a_0^3}{C_M}K_2(k)\right]\nonumber \\
&+&\frac{k^2}{a_0^2}\frac{\cosh(Ct)}{\sinh^{7/3}(Ct)}\left[\frac{4}{3C_M}-\frac{1}{Ca_0^3}\right]B_0(k) \ .
\end{eqnarray}
Using the gauge restrictions of \cite{cvlp} 
\begin{eqnarray} \label{constants}
K_1(k) &=& -\frac{k^2C_M}{2a_0^3}A_k \ ,\  K_2(k)=-\frac{3k^2C_M}{2^{7/3}a_0^5C^2}B_k \ ,\\
H_0(k) &=&\frac{3}{2}B_k \ ,\qquad \quad B_0(k)=0 \ ,\nonumber 
\end{eqnarray}
we can eliminate the non-physical (gauge) terms from the solutions. It is shown in \cite{PVCE} that the $K_1(k)$ term in Eq. (\ref{drho}) corresponds to relatively decreasing density perturbations. Calculating the anisotropy power spectrum of the late ISW effect we are interested in the relatively growing modes, thus in the following we keep only the $K_2(k)$ term in the $\Psi$ and $\Phi$ potentials. Applying these restrictions one can rewrite the solution as
\begin{eqnarray}\label{sol}
\Psi&=&-\frac{1}{2^{4/3}}\frac{\cosh(Ct)}{\sinh^{5/3}(Ct)}I(t)B_k \ , \\
\Phi &=& -\Psi \ , \\
V&=&\frac{3k}{4Ca_0}\left[\frac{I(t)}{2^{1/3}\sinh^{4/3}(Ct)}-\frac{\sinh^{1/3}(Ct)}{\cosh(Ct)}\right]B_k \ , \\
\delta\rho&=&\frac{k^2}{2^{1/3}a_0^2}\frac{\cosh(Ct)I(t)}{\sinh^3(Ct)}B_k \ ,
\end{eqnarray}
where
\begin{eqnarray}\label{I}
I(x)&=&\frac{1}{2^{1/6}}\sqrt{3+\sqrt{3}i}\left( E(x)-\frac{3+\sqrt{3}i}{6}F(x)\right) 
-2^{1/3}\frac{\sqrt{x}(1-x)}{\sqrt{x^{3}+1}} \ , \\
x(t)&=&\frac{a(t)}{a_0}=\sinh^{2/3}(Ct) \ , \label{x}
\end{eqnarray}
and $E$, $F$ are the following incomplete elliptic integrals \cite{PVCE}
\begin{eqnarray}\label{ell}
E(x)&=&E\left(\sqrt{\frac{(3+\sqrt{3}i)x}{2(x+1)}}\left\vert\frac{i-\sqrt{3}}{2}\right. \right),\\ 
F(x)&=&F\left(\sqrt{\frac{(3+\sqrt{3}i)x}{2(x+1)}}\left\vert\frac{i-\sqrt{3}}{2}\right. \right).\nonumber
\end{eqnarray}

Throughout this section we have used the comoving time coordinate $t$ in our calculations. 
In order to derive the anisotropy power spectrum it is advisable to introduce the conformal time variable with the definition $ad\eta=dt$. To transform our results we replace $x(t)$ with 
$x(\eta)$, where
\begin{eqnarray}
x(\eta)=-1 +\frac{3+\sqrt{3}i}{3+\sqrt{3}i-2\ {\rm sn}^2\left(\sqrt{\frac{2}{3(3-\sqrt{3}i)}}Ca_0\eta \left\vert \frac{i-\sqrt{3}}{2}\right.\right)} \ ,
\end{eqnarray}
and sn$(\eta,z)$ is the Jacobi elliptic function (cf. Appendix).

\section{The power spectrum of the late ISW effect}
In the previous section we obtained the complete analytic solution of the scalar perturbation quantities in the Newtonian gauge. In possesion of this result we can deduce the anisotropy 
power spectrum of the late ISW effect. The method of the calculation is well known \cite{HS}, 
therefore we do not present it in details, and hereafter we discuss only the main steps and the 
most important equations of the procedure. 

The power in the $\ell$th multipole is denoted by $C_\ell$,
where 
\begin{equation}\label{cl}
{2 \ell + 1 \over 4\pi} C_\ell = {V \over 2\pi^2} \int_0^\infty {d k \over k}
 k^3 {|\Theta_\ell(\eta,k)|^2 \over 2\ell+1} 
\end{equation}
and $\Theta_\ell$ is the multipole decomposition of the fractional temperature fluctuation \cite{HS}.
In the $\Lambda$ dominated universe the perturbation potentials vary with time and photons experience differential redshifts due to the gradient of $\Psi$, which do not yield equal and opposite contributions as the photons enter and exit the potential well (as is the case in the ordinary Sachs-Wolfe effect), and time dilation from $\Phi$. The sum of these contributions along the line of sight is called the ISW effect. The total contribution is
\begin{eqnarray}\label{SW}
{\Theta_\ell(\eta,k) \over 2\ell+1} =
[\Theta_0&+&\Psi](\eta_*,k) j_{\ell}[k(\eta_0 -\eta_*)]\\
&+& \int_{\eta_*}^{\eta_0} [ \Psi^{\prime} - \Phi^{\prime}](\eta,k)
j_{\ell}[k (\eta_0-\eta)] d\eta\ ,\nonumber
\end{eqnarray}
where the first term represents the ordinary, while the second term the integrated Sachs-Wolfe effect.
In Eq. (\ref{SW}) $j_\ell$ is the Bessel function, $\eta_*$ and $\eta_0$ are the values of the conformal time at decoupling and present respectively, and a prime denotes the derivative with respect to $\eta$. In the following we will focus on the contribution of the ISW term to the anisotropy power
spectrum of CMBR. It is usual to divide the ISW effect into early and late parts. The early ISW effect corresponds to the variations in the perturbation potentials which occurs in the time interval between matter-radiation equality and recombination, while the late ISW effect corresponds to the variations caused by the presence of the cosmological constant. In the $\Lambda$ dominated epoch (which occurs recently) for adiabatic perturbations this late ISW term in Eq. (\ref{SW}) can be approximated with the formula
\cite{HuPhD}
\begin{equation}
\int_{\eta_*}^{\eta_0} [ \Psi^{\prime} - \Phi^{\prime}] j_\ell[k(\eta_0-\eta)] d\eta 
\simeq \frac{I_\ell}{k}[\Psi^{\prime}-\Phi^{\prime}]_{\eta=\eta_k}\ ,
\end{equation}
where $\eta_k = \eta_0 -(\ell+1/2)/k$ and the $I_\ell$ integral is given by 
\begin{equation}
I_\ell \equiv \int_0^\infty dx j_\ell(x) = {\sqrt{\pi} \over 2} 
{\Gamma[{1 \over 2}(\ell+1)] \over \Gamma[{1 \over 2}(\ell+2)]}\ .
\end{equation}
Substituting the results above into Eq. (\ref{cl}) the $C_\ell$ momenta of the late ISW effect have the form   
\begin{equation}\label{isw_int}
C_\ell^{ISW} \simeq 2V\left(\frac{\Gamma[(\ell+1)/2]}{\Gamma[(\ell+2)/2]} \right)^2 \int_0^\infty \Psi^{\prime 2}(\eta_k,k)dk\ .
\end{equation}
In $\Lambda$ cosmological models \cite{HS} the perturbation potential $\Psi$ is related to the growth factor $D$ as $\Psi\sim D/a$, where
\begin{equation}\label{D}
D(a)\sim H\int{\frac{da}{(Ha)^3}} \ 
\end{equation}
and $H$ is the Hubble parameter defined as $H=a^{\prime}/a^2$. 
The integral in Eq. (\ref{D}) has been evaluated so far only with numerical treatment in the presence of cosmological constant \cite{HS}. 
With the solution of the perturbation potential $\Psi$,  we can express the growth factor analytically as follows
\begin{equation}
D\sim a\Psi \sim \coth(Ct)I(t)\equiv \left(\frac{1+x^3}{x^3}\right)^{1/2}I(x) \ .
\end{equation}
Inserting the solution of $\Psi$ into Eq. (\ref{isw_int}) the argument of the integral becomes
\begin{eqnarray}
\Psi^{\prime 2}=\frac{25}{9} 2^{-2/3}C^2a_0^2\left[\frac{2^{1/3}}{\sqrt{x(1+x^3)}}-\frac{1}{3x^3}\left(2x^3+5\right)I(x)\right]^2|\Psi(0,k)|^2 \ ,
\end{eqnarray}
where $\Psi(0,k)=-3B_k/10$. 

In order to calculate the $C^{ISW}_\ell$ momenta, we choose a general power law initial spectrum 
${\cal P}(k)= k^3 |\Psi(0,k)|^2 = Bk^{n-1}$, where $B$ is a constant. It is usual to assume the scale invariant Harrison-Zeldovich spectrum with $n=1$. The matter and dark energy densities of the universe are set to be $\Omega_0=0.3$ and $\Omega_{\Lambda}=0.7$ respectively, according to the combined results of the {\it WMAP} and {\it SDSS} surveys. For the Hubble parameter $H_0=h\cdot 100km/s/Mpc$, we use $h=0.7$. 
   
Having in hand the solution of the perturbation potentials we have expressed the argument of the integral in Eq. (\ref{cl}) analytically, although the integration over $k$ has still to be done numerically. After performing the calculations the obtained power spectrum of the late ISW effect is plotted on Fig. 1. These results are in good agreement with earlier calculations, where the explicit analytic form of the $\Psi$ potential has not been known in the presence of cosmological constant \cite{Starobinsky,HS,HW}.
\begin{figure}
\psfrag{x}{$\ell$}
\psfrag{y}{{$\!\!\!\!\!\!\!\!\!\!\!\!\!\!\!\!\!\!\!\!\!\!\!\!\!\!\!\!\!\ell(\ell+1)C_{\ell}/2\pi(\times 10^{-10} )$}}
\psfrag{L}{\kern -3cm $\Omega_0=0.3 ,\  \Omega_{\Lambda}=0.7$}
\centerline{\includegraphics{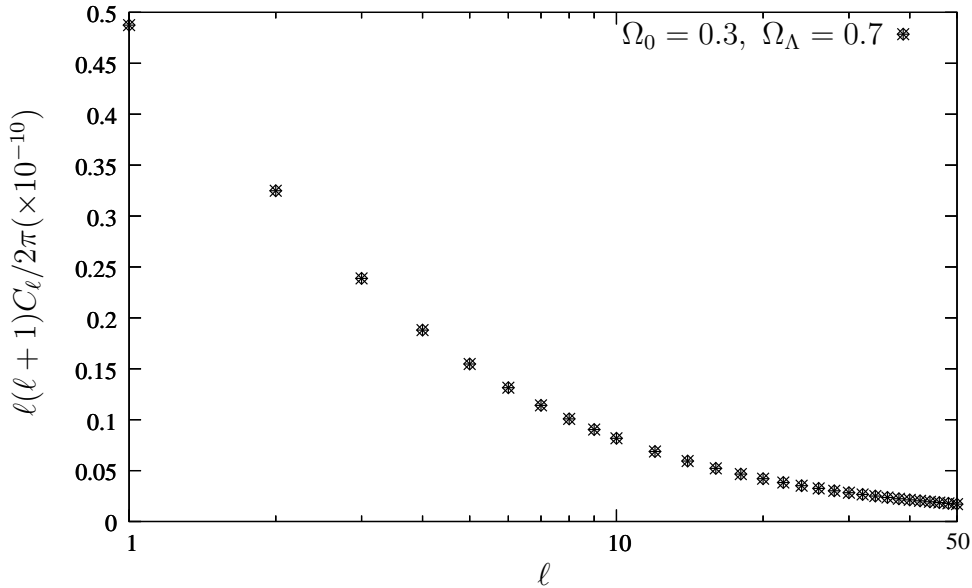}}
\vspace*{8pt}
\caption{The late ISW effect in a $\Lambda$ dominated universe}
\end{figure}

\section{Concluding remarks}
In the present paper we calculated the anisotropy power spectrum of the late ISW effect in a $\Lambda$ dominated universe. We obtained the complete analytic solution for the growth factor and the scalar quantities of the perturbations using the Newtonian gauge in the presence of cosmological constant. 
We also expressed the argument of the ISW integral analytically for an arbitrary initial power spectrum of the fluctuations. We showed the agreement of our results with earlier numerical calculations in the case of the scale invariant Harrison-Zeldovich spectrum.

Our results are relevant in calculating the anisotropy power spectrum in the presence of cosmological constant. In the case of an arbitrary initial spectrum of the fluctuations the anisotropy spectra can be more easily calculated with the help of the obtained analytic formulae.  

\section{Appendix}
The time dependence of the scale factor on the comoving time $t$ is given by Eq. (\ref{backgr}) and Eq. (\ref{x}) as
\begin{equation}
a=a_0\sinh^{2/3}(Ct)\equiv a_0x \ . 
\end{equation}
Using the definition of the conformal time 
\begin{equation}
ad\eta=dt, 
\end{equation}
$\eta$ can be expressed with the following integral of the  variable $x$
\begin{equation}\label{eta}
\eta=\frac{1}{a_0}\int\frac{dt}{x}=\frac{3}{2Ca_0}\int\frac{dx}{\sqrt{x(1+x^3)}} \ .
\end{equation}
Eq. (\ref{eta}) can be integrated analytically with the solution
\begin{equation}
\eta(x)=\sqrt{\frac{3}{2}}\frac{\sqrt{3-\sqrt{3}i}}{Ca_0}F\left(\sqrt{\frac{(3+\sqrt{3}i)x}{2(x+1)}}\left\vert\frac{i-\sqrt{3}}{2}\right. \right) \ , 
\end{equation}
where $F$ is the same incomplete elliptic integral given in (\ref{ell}). Taking its inverse, one can obtain the conformal time dependence of $x$ as
\begin{eqnarray}
x(\eta)=-1 +\frac{3+\sqrt{3}i}{3+\sqrt{3}i-2\ {\rm sn}^2\left(\sqrt{\frac{2}{3(3-\sqrt{3}i)}}Ca_0\eta \left\vert \frac{i-\sqrt{3}}{2}\right.\right)} \ ,
\end{eqnarray}
with the Jacobi elliptic function sn$(\eta,z)$. In spite of the fact that $\eta(x)$ and $x(\eta)$ have complex arguments, both functions are real valued.

\section{Acknowledgments}

This work was supported by OTKA grants no. TS044665, F049429 and T046939.


\begin{thebibliography}{99}

\bibitem{SW} Sachs, R. K., \& Wolfe, A. M. 1967, \apj, 147, 73 

\bibitem{Starobinsky} Kofman, L. A. and Starobinsky, A. A., 1985, 
Sov. Astron. Lett., 11, 271  

\bibitem{Spergel} Spergel, D. N. et al. 2003, \apjs, 148, 175

\bibitem{Tonry} Tonry, J. L. et al. 2003, \apj, 594, 1

\bibitem{Tegmark} Tegmark, M. et al. 2004a, \prd, 69, 103501

\bibitem{HS} Hu, W., \& Sugiyama, N. 1995,\prd, 51, 2599

\bibitem{HW} Hu, W. and White, M., \aanda, 315, 33, 1995

\bibitem{cvlp} Czinner, V., Vas\'uth, M., Luk\'acs, \'A. and Perj\'es, Z.,\\
accepted for publication in IJMPA, [gr-qc/0501009]

\bibitem{Stephani} Stephani, H., Kramer, D., MacCallum, M. A. H., Hoenselaers, C., \& Herlt, E. 2003, \\
Exact Solutions to Einstein's Field Equations, (Cambridge University Press, Cambridge, pp. 211)

\bibitem{Bardeen} Bardeen, J. M. 1980, \prd, 22, 1882

\bibitem{PVCE} Perj\'es Z., Vas\'uth M., Czinner V., and Eriksson D., \\
accepted for publication in \aanda, [astro-ph/0402069]

\bibitem{HuPhD} Hu W., Ph.D. Thesis, 1995

\end{thebibliography}
\end{document}